\documentclass[12pt]{article}
\usepackage{graphicx}
\usepackage{lscape}
\usepackage{citesort}
\begin{document}

\def\ds{\displaystyle}
\def\beq{\begin{equation}}
\def\eeq{\end{equation}}
\def\bea{\begin{eqnarray}}
\def\eea{\end{eqnarray}}
\def\beeq{\begin{eqnarray}}
\def\eeeq{\end{eqnarray}}
\def\ve{\vert}
\def\vel{\left|}
\def\ver{\right|}
\def\nnb{\nonumber}
\def\ga{\left(}
\def\dr{\right)}
\def\aga{\left\{}
\def\adr{\right\}}
\def\lla{\left<}
\def\rra{\right>}
\def\rar{\rightarrow}
\def\nnb{\nonumber}
\def\la{\langle}
\def\ra{\rangle}
\def\ba{\begin{array}}
\def\ea{\end{array}}
\def\tr{\mbox{Tr}}
\def\ssp{{\Sigma^{*+}}}
\def\sso{{\Sigma^{*0}}}
\def\ssm{{\Sigma^{*-}}}
\def\xis0{{\Xi^{*0}}}
\def\xism{{\Xi^{*-}}}
\def\qs{\la \bar s s \ra}
\def\qu{\la \bar u u \ra}
\def\qd{\la \bar d d \ra}
\def\qq{\la \bar q q \ra}
\def\gGgG{\la g^2 G^2 \ra}
\def\q{\gamma_5 \not\!q}
\def\x{\gamma_5 \not\!x}
\def\g5{\gamma_5}
\def\sb{S_Q^{cf}}
\def\sd{S_d^{be}}
\def\su{S_u^{ad}}
\def\ss{S_s^{??}}
\def\sbp{{S}_Q^{'cf}}
\def\sdp{{S}_d^{'be}}
\def\sup{{S}_u^{'ad}}
\def\ssp{{S}_s^{'??}}
\def\sig{\sigma_{\mu \nu} \gamma_5 p^\mu q^\nu}
\def\fo{f_0(\frac{s_0}{M^2})}
\def\ffi{f_1(\frac{s_0}{M^2})}
\def\fii{f_2(\frac{s_0}{M^2})}
\def\O{{\cal O}}
\def\sl{{\Sigma^0 \Lambda}}
\def\es{\!\!\! &=& \!\!\!}
\def\ap{\!\!\! &\approx& \!\!\!}
\def\ar{&+& \!\!\!}
\def\ek{&-& \!\!\!}
\def\kek{\!\!\!&-& \!\!\!}
\def\cp{&\times& \!\!\!}
\def\se{\!\!\! &\simeq& \!\!\!}
\def\eqv{&\equiv& \!\!\!}
\def\kpm{&\pm& \!\!\!}
\def\kmp{&\mp& \!\!\!}

% .........................................................

\def\simlt{\stackrel{<}{{}_\sim}}
\def\simgt{\stackrel{>}{{}_\sim}}

% .........................................................

\title{
         {\Large
                 {\bf
Analysis of exclusive $B_s \rar D_{s_0} (2317) \ell \bar{\nu}_\ell$ decay
in ``full'' QCD
                 }
         }
      }

\author{\vspace{1cm}\\
{\small T. M. Aliev \thanks
{e-mail: taliev@metu.edu.tr}~\footnote{permanent address:Institute
of Physics,Baku,Azerbaijan}\,\,,
M. Savc{\i} \thanks
{e-mail: savci@metu.edu.tr}} \\
{\small Physics Department, Middle East Technical University,
06531 Ankara, Turkey} }

\date{}

\begin{titlepage}
\maketitle
\thispagestyle{empty}

\begin{abstract}
The transition form factors of the semileptonic decay $B_s$ into scalar
$D_{s_0}(2317)$ meson is calculated in the framework of 3--point QCD sum
rule. The branching ratio is found to be $\sim 10^{-3}$  for the
$ B_s \rar D_{s_0}(2317) \ell \bar{\nu}_\ell~(\ell=e,\mu)$ decay, and 
$\sim 10^{-4}$ for the $ B_s \rar D_{s_0}(2317) \tau \bar{\nu}_\tau$ 
decay, respectively.
\end{abstract}

%\vspace{1cm}
%~~~PACS numbers: 12.60.--i, 13.30.--a. 13.88.+e
\end{titlepage}

\section{Introduction}

Semileptonic decays of mesons containing charm and beauty quarks constitute
a very important class of decays for determination of elements of the
Cabibbo-Kobayashi-Maskawa (CKM) matrix, leptonic decay constants of heavy
mesons, as well as for understanding the origin of CP violation which is
related to the structure of the CKM matrix in the Standard Model (SM),
because strong interactions involving semileptonic decays are more simple
compared to that of hadronic decays. In semileptonic decays the long
distance strong binding dynamics can be parametrized as transition
form factors, calculation of which is the main problem of these decays. For
estimation of transition form factors some nonperturbative approach is
needed. Several methods, such as the quark model, lattice QCD, QCD sum
rules, large energy and effective heavy quark theories, have been used to
calculate the transition form factors. Among these approaches the QCD sum
rules occupy a special place, since it is based on the fundamental QCD
Lagrangian.

The QCD sum rules method \cite{R7501} has been successfully applied to a wide
variety of problems in hadron physics (see \cite{R7502,R7503} and references
therein). The semileptonic decay $D \rar \bar{K}^0 e \bar{\nu}_e$ is
first studied in QCD sum rules with the three--point correlation function in
\cite{R7504}. Following this work, $D^+ \rar K^0 e^+ \nu_e$, $D^+ \rar
K^{0\ast} e^+ \nu_e$ \cite{R7505}, $D \rar \pi e \bar{\nu}_e$ \cite{R7506}, $D
\rar \rho e \bar{\nu}_e$ \cite{R7507} and $B \rar D(D^\ast)\ell \bar{\nu}_\ell$
\cite{R7508}, $D \rar \phi \ell \bar{\nu}_\ell$ \cite{R7509} are studied in 
the frame work of the same method. Note that $D_s \rar \phi \ell
\bar{\nu}_\ell$ decay is studied in light cone QCD sum rules \cite{R7510},
which is an alternative approach to the traditional QCD sum rules.     

Transition form factors appearing in semileptonic decays depend not only
on the dynamics of strong interactions between quarks in the initial and
final state hadrons, but also on the structure of the hadrons involved in
the semileptonic decays. Since in the present work we will consider a scalar
meson in the final state, a few words about the scalar mesons are in order.
The structure of the scalar mesons is still under debate. At present, there
are different proposals about the nature of the scalar mesons that have been
put forward, for example, their structures are considered as composed of
$\bar{q}q$, multiquark $\bar{q}q\bar{q}q$ or meson--meson bound states. 

For studying the structure of scalar mesons much more experimental data and
theoretical analysis are needed. The observation of two narrow resonances
with charm and strangeness $D_{s_0}(2317)$ in the $D_s \pi$ invariant mass
distribution \cite{R7511,R7512,R7513,R7514,R7515,R7516,R7517} and
$D_{s_J}(2460)$ in the $D_s^\ast \pi^0$ and $D_s \gamma$ mass distributions
\cite{R7512,R7513,R7514,R7517,R7518,R7519} have raised discussion about the
nature and quark content on these states \cite{R7520}. The natural
identification consists in considering these states as the scalar
$D_{s_0}(2317)$ and axial vector $D_{s_J}(2460)$ $\bar{c}s$ mesons,
respectively. The result of the analysis
of the radiative decays $D_{s_0} (2317) \rar D_s^\ast \gamma$,
$D_{s_J}^\ast (2460) \rar D_s^\ast \gamma$ and $D_{s_J}^\ast (2460) \rar
D_{s_0} (2317) \gamma$ is in favor of the interpretation of 
quark content of these mesons as being ordinary $\bar{c}s$ mesons
\cite{R7521}. It is further observed that deviation of the transition 
amplitude in the infinite heavy quark limit compared to that of the finite 
quark mass case is quite sizeable. In the light of the $D_{s_0} 
\rar D_s^\ast \gamma$ case, one can ask whether a change in the heavy quark
effective theory results occurs for the $B_s \rar D_{s_0} (2317) \ell 
\bar{\nu}_\ell$ decay for the finite quark mass, or not. In the present we
try to answer this question. Note that this decay has been studied within the frame work of the 
heavy quark effective theory in \cite{R7522}.

In this work, we will study the semileptonic decay of $B_s$ meson to
$D_{s_0} (2317)$ meson, i.e., $B_s \rar D_{s_0} (2317) \ell
\bar{\nu}_\ell$, in the frame work of QCD sum rules method. The paper is
organized as follows: In section 2, we derive the sum rules for the
transition form factors. Section 3 is devoted to the numerical analysis and
discussion, and contains a summary of our results and conclusions.        

\section{Sum rules for the $B_s \rar D_{s_0} (2317)$ transition form
factors}

The amplitude of the $B_s \rar D_{s_0} (2317) \ell \bar{\nu}_\ell$ decay
can be presented in the following form
\bea
\label{e7501}
{\cal M} = \frac{G_F}{\sqrt{2}} V_{cb} \bar{\nu}_\ell \gamma_\mu (1-\gamma_5)
\ell \lla D_{s_0} \vel \bar{c} \gamma^\mu (1-\gamma_5) b \ver B_s \rra~,
\eea
where $V_{cb}$ is the CKM matrix element which
describes the transition of a $b$ quark into a $c$ quark, and for the sake
of simplicity, in all further calculations we will denote
$D_{s_0}(2317)$ as $D_0$. The main problem
related to the calculation of the matrix element $\lla D_0 \vel
\bar{c} \gamma^\mu (1-\gamma_5) b \ver B_s \rra$. Obviously, the vector part
of weak $\bar{c} \gamma^\mu (1-\gamma_5) b$ current does not contribute to
the matrix element considered above, which immediately follows from parity
property of the hadrons and weak current; and only axial part of the weak
current gives nonzero contribution. From the Lorentz invariance, this matrix
element can be parametrized in terms of the form factors in the following
way:
\bea
\label{e7502}
\lla D_0 (p^\prime) \vel \bar{c} \gamma_\mu \gamma_5 b \ver
B_s (p) \rra = i \Big[ f_+ {\cal P}_\mu + f_- q_\mu \Big]~,
\eea
where $f_+(q^2)$ and $f_-(q^2)$ are the transition form factors, ${\cal
P}_\mu = (p+p^\prime)_\mu$ and $q_\mu = (p-p^\prime)_\mu$. It is well known
that $f_-$ is proportional to the lepton mass, and especially for the $\tau$
case it can substantially be important. For this reason we will take both
of the form factors into consideration. In calculation of the form
factors $f_+(q^2)$ and $f_-(q^2)$ we will employ the QCD sum rules method
and proceed by considering the following correlator:
\bea
\label{e7503}
\Pi_\mu (p^2,p^{\prime 2},q^2) = i^2 \int d^4x d^4y e^{i(p^\prime y - px)}
\lla 0 \vel J_{D_0} (y) J_\mu^A (0) J_5 (x) \ver 0 \rra~,
\eea
where $J_{D_0} (y) = \bar{s}c$, $J_5 = \bar{s} \gamma_5 b$ and 
$J_\mu^A = \bar{c} \gamma_\mu \gamma_5 b$ are the interpolating currents of 
the scalar $D_0$, $B_s$ mesons and weak axial currents, respectively. 

Let us first calculate the phenomenological part of the correlator given in
Eq. (\ref{e7503}). This can be obtained by inserting the complete set of
intermediate states with the same quantum number as the currents $J_{D_0}$
and $J_5$. isolating the pole terms of the lowest scalar and pseudoscalar
$D_0$ and $B_s$ mesons, we get the following representation of
the above--mentioned correlator    
\bea
\label{e7504}
\Pi_\mu  (p^2,p^{\prime 2},q^2) \es \frac{\ds \lla 0 \vel J_{D_0} \ver
D_0 \rra \lla D_0 \vel \bar{c} \gamma_\mu \gamma_5 b
\ver B_s \rra \lla B_s \vel \bar{s} \gamma_5 b \ver 0 \rra}
{\ds (m_{D_0}^2 - p^{\prime 2}) (m_{B_s}^2 - p^2)} \nnb \\
\ar \sum_h \frac{\ds \lla 0 \vel J_{D_0} \ver
h(p^\prime) \rra \lla h(p^\prime) \vel \bar{c} \gamma_\mu \gamma_5 b  
\ver {\cal H}(p) \rra \lla {\cal H}(p) \vel \bar{s} \gamma_5 b \ver 0 \rra}
{\ds (p^{\prime 2} - m_h^2) (p^2-m_{ {\cal H}}^2)}~.
\eea 
The second term in Eq. (\ref{e7504}) takes into account higher states and
continuum contributions; and $h$ and ${\cal H}$ form a complete set of
mesons having the same quantum numbers as the ground state mesons.

The matrix elements in Eq. (\ref{e7504}) are defined in the standard way as:
\bea
\label{e7505}
\lla 0 \vel J_{D_0} \ver D_0 \rra \es f_{D_0}
m_{D_0} ~, \nnb \\
\lla B_s \vel \bar{s} \gamma_5 b \ver 0 \rra \es -i \frac{\ds 
f_{B_s} m_{B_s}^2}{\ds m_b+m_s}~,
\eea
where $f_{D_0}$ and $f_{B_s}$ are the leptonic decay constants
of $D_0$ and $B_s$ mesons, respectively. Using
(\ref{e7505}), Eq. (\ref{e7504}) can be written as
\bea
\label{e7506}
\Pi_\mu  (p^2,p^{\prime 2},q^2) \es - \frac{\ds  f_{B_s} m_{B_s}^2}{\ds
(m_b+m_s)}     
\frac{\ds f_{D_0} m_{D_0}}{\ds (m_{D_0}^2 - p^{\prime 2}) (m_{B_s}^2 - p^2)}
\Big[ f_+ {\cal P}_\mu + f_- q_\mu \Big] + \mbox{\rm excited states}~.
\eea
In accordance with the QCD sum rules philosophy, $\Pi_\mu  (p^2,p^{\prime
2},q^2)$ can be calculated from QCD side wit the help of the operator
product expansion method (OPE) in the deep Euclidean region $p^2 \ll
(m_b+m_c)^2$ and $p^{\prime 2} \ll (m_c+m_s)^2$. Equating the two different
representations of $\Pi_\mu$ gives us sum rules for the form factors 
$f_+(q^2)$ and $f_-(q^2)$. 

The theoretical part of the correlator is calculated by means of OPE, and
up to operators having dimension d=6, it is determined by the bare--loop and
the power corrections (Figs. (1)--(3)), from the operators with d=3
$\la \bar{\psi}\psi \ra$, d=5 $m_0^2 \la \bar{\psi}\psi \ra$
and d=6 $\la \bar{\psi}\psi \ra^2$. Our calculation shows that
d=4 operator $\lla G_{\mu\nu}^2 \rra$ gives very small contribution, and for
this reason we do not consider it in the present work. 

In calculating the bare--loop contribution, we we first write the double
dispersion representation as
\bea
\label{e7507}
f_i^{per} = - \frac{1}{(2 \pi)^2} \int ds ds^\prime
\frac{\rho_i(s,s^\prime,q^2)}{(s-p^2) (s^\prime-p^{\prime 2})} +
\mbox{\rm subtraction terms}~.
\eea  

The spectral density $\rho_i(s,s^\prime,q^2)$ can be calculated from the
usual Feynman integral with the help of Gutkovsky rule, i.e., by replacing
the denominators of the propagators as follows:
\bea
\frac{1}{p^2-m^2} \rar - 2 \pi i \delta(p^2-m^2)~,\nnb
\eea
which implies that all quarks are real.

After standard calculations for the spectral densities we obtain:
\bea
\label{e7508}
\rho_+(s,s^\prime,q^2) \es \frac{1}{4 \lambda^{1/2}(s,s^\prime,q^2)} 
\Big\{ (\Delta^\prime + \Delta) (1+A+B) +
[(m_b+m_c)^2 - q^2 ] (A+B) \nnb \\
\ar 2 m_s (m_b-m_c) (1+A+B) \Big\}~ \nnb \\ \nnb \\
\rho_-(s,s^\prime,q^2) \es \frac{1}{4 \lambda^{1/2}(s,s^\prime,q^2)} 
\Big\{ [\Delta^\prime + \Delta +
(m_b+m_c)^2 - q^2 + 2 m_s (m_b-m_c)] (A-B) \nnb \\
\ar \Delta^\prime - \Delta - 2 m_s (m_b+m_c) \Big\}~,
\eea
where $\Delta^\prime = s-m_c^2$ and $\Delta=s-m_b^2$, and
\bea
A \es \frac{1}{\lambda (s,s^\prime,q^2)} [ -(s+s^\prime -q^2) \Delta^\prime
+ 2 \Delta s^\prime ]~, \nnb \\
B \es \frac{1}{\lambda (s,s^\prime,q^2)} [ -(s+s^\prime -q^2) \Delta 
+ 2 \Delta^\prime s ]~, \nnb
\eea
and $\lambda (s,s^\prime,q^2)=s^2 + s^{\prime 2} +q^4 -2 s q^2 -2 s^\prime
q^2 -2 s s^\prime$. Here and in all following expressions, subscripts $+$ and
$-$ correspond to the coefficients of the structures proportional to ${\cal
P}_\mu$ and $q_\mu$, respectively. Note that, in deriving Eqs. (\ref{e7507})
and (\ref{e7508}) we retain only linear terms in $m_s$ in order to take
SU(3) violation effects into account, and higher
order $m_s$ terms are all neglected. The integration region in
for the perturbative contribution in Eq. (\ref{e7507}) is determined from
the condition that arguments of the three $\delta$ functions might vanish
simultaneously. The physical region in $s$ and $s^\prime$ plane is described
by the following inequalities:
\bea
-1 \le \frac{\ds 2 s s^\prime - 2 m_c^2 s + (s+s^\prime - q^2) (m_b^2-s)}
{\lambda^{1/2}(s,s^\prime,q^2) (m_b^2-s)} \le +1~. \nnb
\eea
According to the quark--hadron duality, the contribution of higher states in
phenomenological part is parametrized in correspondence with the spectral
density starting from $s>s_0$ and $s^\prime >s_0^\prime$.

In what follows, we present the contributions of d=3, 5 and 6 operators.
\bea
\label{e7509}
f_+^{(3)} \es \frac{1}{2} \lla \bar{s}s \rra \frac{m_b-m_c}{r r^\prime}~, \\
\label{e7510}
f_-^{(3)} \es -\frac{1}{2} \lla \bar{s}s \rra \frac{m_b+m_c}{r r^\prime}~, \\
\label{e7511}
f_+^{(4)} \es \frac{1}{4} m_s \lla \bar{s}s \rra \Bigg[ - \frac{m_c(m_b-m_c)}
{r r^{\prime 2}} + \frac{m_b(m_b-m_c)}{r^2 r^\prime} - \frac{2}{r r^\prime}
\Bigg]~, \\
\label{e7512}
f_-^{(4)} \es \frac{1}{4} m_s (m_b+m_c) \lla \bar{s}s \rra \Bigg[ \frac{m_c}
{rr^{\prime 2}} - \frac{m_b}{r^2 r^\prime} \Bigg]~, \\
\label{e7513}
f_+^{(5)} \es - \frac{1}{12} m_0^2 \lla \bar{s}s \rra \Bigg[
\frac{3 m_c^2 (m_b-m_c)}{r r^{\prime 3}} + \frac{3 m_b^2 (m_b-m_c)}{r^3
r^\prime} + \frac{2 (m_b-2 m_c)}{r r^{\prime 2}} \nnb \\
\ar \frac{2 (2 m_b-m_c)}{r^2 r^\prime}
+ \frac{(m_b-m_c)(2 m_b^2+m_b m_c +2 m_c^2-2 q^2)}{r^2 r^{\prime 2}} \Bigg]~, \\
\label{e7514}
f_-^{(5)} \es \frac{1}{12} m_0^2 \lla \bar{s}s \rra \Bigg[ 
\frac{3 m_c^2 (m_b+m_c)}{r r^{\prime 3}} + \frac{3 m_b^2 (m_b+m_c)}
{r^3 r^\prime} + \frac{2 (m_b+3 m_c)}{r r^{\prime 2}} \nnb \\
\ar \frac{2 (m_c+3 m_b)}{r^2 r^\prime}        
+ \frac{(m_b+m_c)(2 m_b^2+m_b m_c + 2 m_c^2-2 q^2)}{r^2 r^{\prime 2}}
\Bigg]~, \\
\label{e7515}
f_+^{(6)} \es \frac{4}{81} \pi \alpha_s \lla \bar{s}s \rra^2 \Bigg[ 
\frac{12 m_c^3 (m_b-m_c)}{r r^{\prime 4}} - \frac{12 m_b^3 (m_b-m_c)}
{r^4 r^\prime} \nnb \\
\ar \frac{4 m_c (m_b-m_c)(2 m_b^2+m_b m_c + 2 m_c^2-2 q^2)}{r^2 r^{\prime 3}}
+ \frac{8 m_c (7 m_b-8 m_c)}{r r^{\prime 3}} \nnb \\
\ek \frac{4 m_b (m_b-m_c)(2 m_b^2+m_b m_c + 2 m_c^2-2 q^2)}{r^3 r^{\prime 2}}
- \frac{8 m_b (8 m_b-7 m_c)}{r^3 r^\prime} \nnb \\
\ek \frac{4 (5 m_b^2-20 m_b m_c + 5 m_c^2+2 q^2)}{r^2 r^{\prime 2}} +
\frac{48}{r r^{\prime 2}} + \frac{48}{r^2 r^\prime}\Bigg] \nnb \\
\ar \frac{1}{9} m_0^2 m_s \lla \bar{s}s \rra \Bigg[
- \frac{6 m_c^3 (m_b-m_c)}{r r^{\prime 4}} - 
\frac{m_c (m_b-m_c)(5 m_b^2+4 m_b m_c + 5 m_c^2-5 q^2)}{r^2 r^{\prime 3}}
\nnb \\
\ek \frac{2 m_c (7 m_b-11 m_c)}{r r^{\prime 3}} -
\frac{m_b (m_b-m_c)(m_b^2+8 m_b m_c + m_c^2- q^2)}{r^3 r^{\prime 2}} \nnb \\
\ar \frac{5 m_b^2- 20 m_b m_c + 11 m_c^2-4 q^2}{r^2 r^{\prime 2}} +
\frac{6 m_b^3 (m_b-m_c)}{r^4 r^\prime} +
\frac{2 m_b (5 m_b-13 m_c)}{r^3 r^\prime} \Bigg]~, \\
\label{e7516}
f_-^{(6)} \es \frac{4}{81} \pi \alpha_s \lla \bar{s}s \rra^2 \Bigg[ 
\frac{- 12 m_c^3 (m_b+m_c)}{r r^{\prime 4}} + \frac{12 m_b^3 (m_b+m_c)}
{r^4 r^\prime} \nnb \\
\ek \frac{4 m_c (m_b+m_c)(2 m_b^2+m_b m_c + 2 m_c^2-2 q^2)}{r^2 r^{\prime 3}}
- \frac{8 m_c (7 m_b+9 m_c)}{r r^{\prime 3}} \nnb \\
\ar \frac{4 m_b (m_b+m_c)(2 m_b^2+m_b m_c + 2 m_c^2-2 q^2)}{r^3 r^{\prime 2}}
+ \frac{8 m_b (9 m_b+7 m_c)}{r^3 r^\prime} \nnb \\
\ar \frac{28 (m_b^2-m_c^2)}{r^2 r^{\prime 2}} +
\frac{8}{r r^{\prime 2}} - \frac{8}{r^2 r^\prime}\Bigg] \nnb \\
\ar \frac{1}{9} m_0^2 m_s \lla \bar{s}s \rra \Bigg[
\frac{6 m_c^3 (m_b+m_c)}{r r^{\prime 4}} +
\frac{m_c (m_b+m_c)(5 m_b^2+4 m_b m_c + 5 m_c^2-5 q^2)}{r^2 r^{\prime 3}}
\nnb \\
\ar \frac{2 m_c (7 m_b+12 m_c)}{r r^{\prime 3}} +
\frac{m_b (m_b+m_c)(m_b^2+8 m_b m_c + m_c^2- q^2)}{r^3 r^{\prime 2}} \nnb \\
\ek \frac{m_b^2- 18 m_b m_c - 7 m_c^2}{r^2 r^{\prime 2}} -
\frac{6 m_b^3 (m_b+m_c)}{r^4 r^\prime} +
\frac{4}{r r^{\prime 2}} - \frac{4}{r^2 r^\prime} 
- \frac{2 m_b (12 m_b+13 m_c)}{r^3 r^\prime} \Bigg]~,
\eea
where $r=p^2-m_b^2$, $r^\prime=p^{\prime 2}-m_c^2$.
Note that, in the present work we neglect the ${\cal O}(\alpha_s)$ corrections
to the bare loop. For consistency, we also neglect ${\cal O}(\alpha_s)$
corrections in determination of the leptonic decay constants $f_{B_s}$ and
$f_{D_0}$.

Substitute Eqs. (\ref{e7508})--(\ref{e7516}) into Eq. (\ref{e7507}) and
applying double Borel transformation $\hat{{\cal B}}$ with respect to the 
variables $p^2$ and $p^{\prime 2}$ $(p^2 \rar M_1^2,~p^{\prime 2} \rar
M_2^2)$ in order to suppress the contributions of higher states and
continuum, we get the following sum rules for the form factors $f_+$ and
$f_-$:
\bea
\label{e7517}
f_\pm(q^2)  \es - \frac{(m_b+m_c)}{f_{B_s} m_{B_s}^2} \frac{1}{f_{D_s}
m_{D_S}} e^{ (m_{B_s}^2/M_1^2 + m_{D_s}^2/M_2^2) } \Bigg[
\hat{B} \Big( f_\pm^{(3)} + f_\pm^{(4)} + f_\pm^{(5)} + f_\pm^{(6)}
\Big) \nnb \\
\ek \frac{1}{(2 \pi)^2} \int ds ds^\prime \rho_\pm (s,s^\prime,q^2) 
 e^{ - s/M_1^2 - s^\prime/M_2^2 } \Bigg]~.
\eea
Double Borel transformation is implemented by the following expression:
\bea
\hat{{\cal B}} \frac{1}{r^m} \frac{1}{r^{\prime n}} \rar (-1)^{m+n}
\frac{1}{\Gamma(m)} \frac{1}{\Gamma(n)} e^{-s/m_b^2}
e^{-s^\prime/m_c^2}~.\nnb 
\eea

\section{Numerical analysis}

In this section we present our numerical analysis for the form factors
$f_+(q^2)$ and $f_-(q^2)$. It follows from the expressions of these form 
factors that the main input parameters needed are the condensates, leptonic
decay constants of $B_s$ and $D_0$ mesons, continuum thresholds $s_0$
and $s_0^\prime$ and Borel parameters $M_1^2$ and $M_2^2$.

In further numerical analysis we choose the value of the condensates at a 
fixed renormalization scale of about $1~GeV$. The values of the condensates 
are taken from \cite{R7522} and can be listed as follows:
\bea
\la \bar{\psi}\psi \ra \Big|_{\mu=1~GeV} \es - (240 \pm 10 ~MeV)^3~,\nnb \\
\la \bar{s}s \ra \es (0.8 \pm 0.1) \la \bar{\psi}\psi \ra~. \nnb
\eea
The quark masses are taken to be $m_c=(\mu=m_c) = 1.275 \pm 0.015~GeV$,
$m_s(1~GeV)\simeq 142~MeV$ \cite{R7522}, $m_b=(4.7 \pm 0.1)~GeV$
\cite{R7523}. For the values of the leptonic decay constants of $B_s$ and
$D_0$ mesons we use the results obtained from two--point QCD analysis:
$f_{B_s}=209 \pm 38~MeV$ \cite{R7503} and $f_{D_0}=225 \pm 25~MeV$
\cite{R7521}. The threshold parameters $s_0$ and $s_0^\prime$ are also
determined from the two point QCD sum rules: $s_0=(35 \pm 2)~GeV^2$
\cite{R7503} and $s_0^\prime=(2.5~GeV)^2$ \cite{R7521}. The Borel parameters
$M_1^2$ and $M_2^2$ are auxiliary quantities and therefore the results of
physical quantities should not depend on them, if OPE can be calculated up to
all order. In QCD sum rule method, OPE is truncated at some finite order.
For this reason, we need to choose "working" regions for the Borel
parameters where form factors are supposed to be  practically independent of
them. The choice of the working region for the Borel parameters $M_1^2$ and
$M_2^2$ should be based, on the one side, on the condition that that the
continuum contribution should be small, and on the other side, the
convergence of the power corrections. As a result of the above--mentioned
conditions, the best stability is achieved for $10~GeV^2 \le M_1^2 \le
15~GeV^2$ and $4~GeV^2 \le M_2^2 \le 7~GeV^2$.

As a result of the above--summarized considerations, our analysis leads to
th following predictions for the form factors at $q^2=0$:
\bea
\label{e7518}
f_+ \es 0.20 \pm 0.05 ~, \nnb \\   
f_- \es -0.32 \pm 0.08 ~. 
\eea 
The errors are can be attributed to the variation of the thresholds, decay
constants, uncertainties in condensates and in quark masses. 

For completeness, we also present the results of HQET for the
above--mentioned form factors that predicts, $f_+=-0.37$ and $f_-=-0.15$.
Indeed, we observe that finite quark mass effects are essential in
determination of the form factors.

In order to estimate the width of $B_s \rar D_0 \ell \bar{\nu}_\ell$ it is
necessary to know the $q^2$ dependence of the form factors $f_+(q^2)$ and
$f_-(q^2)$ in the whole physical region $m_\ell^2 \le q^2 \le
(m_{B_s}-m_{D_0})^2$. The $q^2$ dependence of the form factors can be
calculated from QCD sum rules (for details, see \cite{R7505,R7506}). In the
present work we have analyzed this dependence and used it in our numerical
calculations.

Having decided on the parametrization of the form factors, it is not difficult 
to obtain the expression for the differential decay rate
\bea
\frac{d\Gamma}{dq^2} \es \frac{1}{192 \pi^3 m_{B_s}^3} G^2 \vel V_{cb}
\ver^2 \lambda^{1/2}(m_{B_s}^2,m_{D_0}^2,q^2) \ga \frac{q^2-m_\ell^2}{q^2}
\dr^2 \nnb \\
\cp \Bigg\{ -\frac{(2 q^2+m_\ell^2)}{2}  \Big[ \vel f_+(q^2) \ver^2 (2
m_{B_s}^2 + 2 m_{D_0}^2 - q^2 ) + 2 (m_{B_s}^2 - m_{D_0}^2)
\mbox{\rm Re} [f_+(q^2) f_-^\ast(q^2)] \nnb \\
\ar \vel f_-(q^2) \ver^2 q^2 \Big]
+ \frac{(q^2+ 2 m_\ell^2)}{q^2} \Big[ \vel f_+(q^2) \ver^2 (m_{B_s}^2 -
m_{D_0}^2)^2 \nnb \\
\ar 2 (m_{B_s}^2 - m_{D_0}^2) q^2 
\mbox{\rm Re} [f_+(q^2) f_-^\ast(q^2)] + \vel f_-(q^2) \ver^2 q^4 \Big] \Bigg\}~. \nnb
\eea
Taking into account the $q^2$ dependence of the form factors $f_+$ and
$f_-$, and performing integration over $q^2$ and using the total life--time 
$\tau_{B_s} = 1.46 \times 10^{-12}~s$ \cite{R7524}, we get for the branching ratio
\bea
{\cal B} (B_s \rar D_0 \ell \bar{\nu}_\ell) \se 10^{-3}~,~~~
(\ell = e,~\mu)~, \nnb \\
{\cal B} (B_s \rar D_0 \tau \bar{\nu}_\tau) \se 10^{-4}~. \nnb
\eea
Our predictions on the branching ratio is considerably smaller compared to
that of the HQET results. 

At the end of our analysis, we would like to note that, although current $B$
meson factories do not produce $B_s$ meson, it is hoped to be possible to
study the weak decays of $B_s$ mesons in future--planned experiments at LHC.
The exclusive $B_s \rar D_0 \ell \bar{\nu}_\ell$ decay can be studied via the
resonant production of the scalar $D_0$ meson in the weak decay of the
$B_s$ meson. This analysis can give valuable essential information about the
quark content of the scalar $D_0$ meson.

In summary, we study the semileptonic $B_s \rar D_0 \ell \bar{\nu}_\ell$
decay in the framework of 3--point QCD sum rules. We calculate the
transition form factors, and using these predictions, we estimate
the branching ratio of the $B_s \rar D_0 \ell \bar{\nu}_\ell$ decay.

\end{document}